\documentclass[12pt]{article}
\usepackage{amssymb,amsmath,epsfig}
\allowdisplaybreaks

\begin{document}

\title{\bf Noncommutative Wormhole Solutions in Einstein Gauss-Bonnet Gravity}
\author{Shamaila Rani \thanks {drshamailarani@ciitlahore.edu.pk} and
Abdul Jawad \thanks {abduljawad@ciitlahore.edu.pk, jawadab181@yahoo.com}\\
Department of Mathematics, COMSATS Institute of\\ Information
Technology, Lahore-54000, Pakistan.}
\date{}

\maketitle
\begin{abstract}
In this paper, we explore static spherically symmetric wormhole
solutions in the framework of $n$-dimensional Einstein Gauss-Bonnet
gravity. Our objective is to find out wormhole solutions that
satisfy energy conditions. For this purpose, we consider two
frameworks such as Gaussian distributed and Lorentzian distributed
non-commutative geometry. Taking into account constant redshift
function, we obtain solutions in the form of shape function. The
fifth and sixth dimensional solutions with positive as well as
negative Gauss-Bonnet coefficient are discussed. Also, we check the
equilibrium condition for the wormhole solutions with the help of
generalized Tolman-Oppenheimer-Volkov equation. It is interesting to
mention here that we obtain fifth dimensional stable wormhole
solutions in both distributions that satisfy the energy conditions.
\end{abstract}

\section{Introduction}

The study of wormhole solution become a prime focus of interest in
the modern cosmology as it connects different distant parts of the
universe as a shortcut. The wormhole is like a tunnel or bridge with
two ends which are open in distant parts of the universe to join. To
develop the mathematical structure of wormhole in general relativity
\cite{9}, the basic ingredient is an energy-momentum tensor which
constitute exotic matter. This is hypothetical form of matter
results the violation of energy conditions. For two way travel, the
traversable condition must fulfill, i.e., the throat of wormhole
must remain open due to violation of the null energy condition.
Since normal matter satisfies the energy conditions, so the matter
violating the energy condition is called exotic. The phantom dark
energy violates the energy conditions may be a reasonable source of
wormhole construction \cite{10b,10a}. The inclusion of some scalar
field models, electromagnetic field, Gaussian and Lorentzian
distributions of non-commutative geometry, thin-shell formalism etc
demonstrate more interesting and useful results
\cite{11+f}-\cite{11c}.

In order to minimize the usage of exotic matter or to find another
source of violation while normal matter satisfy the energy
conditions, many directions are adopted so far. Modified theories of
gravity is the most appealing direction which contribute using
effective energy-momentum tensor. For instance, in $f(R)$ \cite{13}
as well as $f(T)$ \cite{5} theories, it has been proved that the
effective energy-momentum tensor which consists on higher order
curvature terms or some torsion terms is responsible for the
necessary violation to traverse through the wormhole. Now-a-days,
higher dimensional wormhole solutions are also under discussion.
Many theories point out the existence of extra dimensions in the
universe which leads to explore wormhole solutions for higher
dimensions.

Rahaman with his collaborators have done a lot of work taking
non-commutative geometry in four dimensional spacetime as well as
higher dimensional cases. Rahaman et al. \cite{1} studied the higher
dimensional static spherically symmetric wormhole solutions in
general relativity  with Gaussian distribution. They found these
solutions upto four dimensions while for fifth dimensional case, in
a very restrictive way. Bhar and Rahaman \cite{2} obtained the same
result by taking Lorentzian distribution. Rahaman et al. \cite{2+}
worked for viable physical properties of new wormhole solutions
inspired by non-commutative geometry with conformal killing vectors.
In $f(R)$ gravity with Gaussian \cite{3} and Lorentzian
distributions \cite{11-e}, wormhole solutions are constructed but
with violation of energy conditions. Sharif and Rani \cite{5}
explored the non-commutative wormhole solutions in $f(T)$ gravity
and found some physically acceptable solutions.

In a recent paper, Jawad and Rani \cite{6} studied Lorentz
distributed wormhole solutions in $f(T)$ gravity and found some
stable wormhole solutions satisfying energy conditions. In the
higher dimensional gravity theories, $n$-dimensional Einstein
Gauss-Bonnet gravity is widely used. The modern string theory
established its natural appearance in the low energy effective
action. Bhawal and Kar \cite{7} studied Lorentzian wormhole
solutions in $D$-dimensional Einstein Gauss-Bonnet gravity which
depend on the dimensionality of the spacetime and coupling
coefficient of Gauss-Bonnet combination. Taking traceless fluid,
Mehdizadeh et al. \cite{8} extended this work and found wormhole
solutions which satisfy energy conditions.

We explore wormhole solutions taking spacetime of $(n-2)$ sphere in
the framework of $n$-dimensional Einstein Gauss-Bonnet gravity. We
consider two frameworks: non-commutative geometry having Gaussian
distributed energy density and Lorentzian distributed energy
density. For $n=5$ and $6$-dimensions, we take positive as well as
negative Gauss-Bonnet coefficient. Also, we check the stability of
wormhole solutions with the help of generalized
Tolman-Oppenheimer-Volkov equation. The paper is organized as
follows: In the next section, we construct the field equation for
higher dimensional wormhole solutions in $n$-dimensional
Gauss-Bonnet gravity. Section \textbf{3} is devoted to the
construction of wormhole solutions in Gaussian and Lorentzian
non-commutative frameworks. In section \textbf{4}, we check the
equilibrium condition of the wormhole solutions. Last section
summarizes the discussions and results.

\section{Field Equations}

In this section, we provide some basic and brief reviews about
$n$-dimensional Einstein Gauss-Bonnet gravity as well as wormhole
geometry and construct field equations in the underlying scenario.

\subsection{$n$-dimensional Einstein Gauss-Bonnet Gravity}

The action for $n$-dimensional Einstein-Gauss-Bonnet gravity is a
outcome of string theory in low energy limit. It is given by
\begin{equation}\label{1}
S_{_{nGB}}=\int d^{n}x\sqrt{-g}[R-\epsilon_1 \mathcal{L}_{_{GB}}],
\end{equation}
where $R$ is the $n$-dimensional Ricci scalar, $\epsilon_1$ is the
Gauss-Bonnet coefficient and $\mathcal{L}_{_{GB}}$ is the
Gauss-Bonnet term defined as
\begin{equation}\label{2}
\mathcal{L}_{_{GB}}=R^2-4R_{\alpha\beta}R^{\alpha\beta}+R_{\alpha\beta\gamma\delta}R^{\alpha\beta\gamma\delta},
\end{equation}
Varying the action with respect to metric tensor, the field
equations become
\begin{equation}\label{3}
G_{\alpha\beta}+\epsilon_1\mathbb{G}_{\alpha\beta}=\mathcal{T}_{\alpha\beta},
\end{equation}
where $G_{\alpha\beta}$ and $\mathcal{T}_{\alpha\beta}$ are the
Einstein and energy-momentum tensors respectively while Gauss-Bonnet
tensor is defined as
\begin{equation}\label{4}
\mathbb{G}_{\alpha\beta}=2(RR_{\alpha\beta}-2R_{\alpha\gamma}{R^\gamma}_{\beta}-2R_{\alpha\gamma\beta\delta}
R^{\gamma\delta}-R_{\alpha\delta\sigma\tau}{R^{\sigma\tau\delta}}_{\beta})-\frac{1}{2}\mathcal{L}_{_{GB}}g_{\alpha\beta}.
\end{equation}
It is noted that we assume $8\pi G_n=1$ where $G_n$ is the
$n$-dimensional gravitational constant.

\subsection{Wormhole Geometry}

The wormhole spacetime for $(n-2)$ sphere is given by \cite{1,2}
\begin{equation}\label{5}
ds^2=-e^{2\lambda(r)}dt^2+\frac{dr^2}{1-\frac{s(r)}{r}}+r^2d\Omega_{n-2}^2,
\end{equation}
where $\lambda(r)$ is the redshift function and $s(r)$ is the shape
function. For traversable wormhole scenario, we have to choose
$\lambda$ to be finite to satisfy the no-horizon condition. Usually,
it is taken as zero for the sake of simplicity, which gives
$e^{2\lambda(r)}\rightarrow 1$. The reason behind finite redshift
function is as follows. The redshift function defines that part of
the metric responsible for finding the magnitude of the
gravitational redshift. The gravitational redshift is the reduction
in the frequency that a photon will experience when it climbs out
from gravitational potential well in order to escape to infinity. In
doing so, the photon uses energy. Its energy is proportional to its
frequency. A reduction in energy, then is equivalent to a reduction
in frequency, which is also known as redshift function. If the
wormhole has an event horizon, it means that a photon emitted
outwardly from the horizon cannot escape to infinity. In other
words, it would take an infinite amount of energy for the photon to
escape. Its frequency would be infinity reduced, i.e., its redshift
would be negatively infinite. A negatively infinite value of the
redshift function at a particular value of the radial coordinate
indicates the presence of an event horizon there. Thus to be
traversable wormhole solution, the magnitude of its redshift
function must be finite.

The shape of the wormhole is such that a spherical hole in space
with increasing length of diameter as moving far from throat (the
minimum non-zero value of radial coordinate denoted as $r_0$) and
combine two asymptomatically flat regions. In order to have a proper
shape of the wormhole, the shape function must attain the ratio to
radial coordinate as $1$ and represents increasing behavior with
respect to radial coordinate which is $1-\frac{s(r)}{r}\geq0$. This
condition of ratio is known as flare-out condition. In addition, the
value of shape function and radial coordinate must be same at
throat, i.e., $s(r_0)=r_0$. There are also some other constraints
applied on the derivative of shape functions which must satisfy.
These are $rs'-s<0$ and $s'(r_0)<1$. Also, the proper distance,
$D(r)=\pm\int^{r}_{r_0}(1-\frac{s(r)}{r})^{-\frac{1}{2}}dr$ must
meet the criteria as decreasing behavior from upper region
$D=+\infty$ towards throat where $d=0$ and then towards lower region
where $D=-\infty$.

In order to make the wormhole to be traversable, the throat must
remain open. To prevent shrinking of wormhole throat, there must
exists such form of energy-momentum tensor which provides the
corresponding matter content. This matter content violates the
energy conditions in order to keep throat open and thus, named as
exotic matter. This implies that violation of these conditions is
the basic key ingredient to construct traversable wormhole
solutions. Since the usual energy-momentum tensor satisfies the
energy conditions. Therefore, the search for wormhole solutions for
which violation may come from other source while matter content
satisfies energy conditions becomes one of the most challenging
problem in astrophysics.

The higher dimensional gravity theories and modified theories may
play positive role by providing violation from higher order
Lagrangian terms and effective form of energy-momentum tensor. The
relationship between Raychaudhuri equation and attractiveness of
gravity yields the \textbf{weak energy condition (WEC)} as
$\mathcal{T}_{\alpha\beta}\mu^\alpha \mu^\beta\geq0$, for any
timelike vector $\mu^\alpha$. In terms of components of the
energy-momentum tensor, this inequality yields $\rho\geq0$ and
$\rho+p\geq0$. The \textbf{null energy condition (NEC)} is developed
by continuity through WEC, i.e., the NEC is
$\mathcal{T}_{\alpha\beta}\chi^\alpha \chi^\beta\geq0$, for any null
vector $\chi^\alpha$. This inequality gives $\rho+p\geq0$. Also, it
is noted that WEC keeps NEC.

The anisotropic energy-momentum is given by
\begin{equation}\label{6}
\mathcal{T}_{\beta}^{\alpha}=(\rho+p_r)u^\alpha u_\beta-p_r
g^\alpha_\beta+(p_t-p_r)\eta^\alpha \eta_\beta,
\end{equation}
where $p_r$ and $p_t$ are the radial and tangential pressure
components with $\rho=\rho(r),~p=p(r)$ and satisfy $u^\alpha
u_\beta=-\eta^\alpha \eta_\beta=1$. Using Eq.(\ref{3}), the field
equations become
\begin{eqnarray}\label{7}
\rho(r)&=&\frac{(n-2)}{2r^2}\left[\left(s'-\frac{s}{r}\right)\left(1+\frac{2\epsilon
s}{r^3}\right)+\frac{s}{r}\left\{(n-3)+(n-5)\frac{\epsilon
s}{r^3})\right\}\right],\\\label{8}
p_r(r)&=&\frac{(n-2)}{2r}\left[2\left(1-\frac{s}{r}\right)\left(1+\frac{2\epsilon
s}{r^3}\right)\lambda'-\frac{s}{r^2}\left\{(n-3)+(n-5)\frac{\epsilon
s}{r^3}\right\}\right],
\\\nonumber p_t(r)&=&\left(1-\frac{s}{r}\right)\left(1+\frac{2\epsilon
s}{r^3}\right)\left[\lambda''+\lambda'^2+\frac{(s-rs')\lambda'}{2r(r-s)}\right]
+\left(1-\frac{s}{r}\right)\left(\frac{\lambda'}{r}\right.\\\nonumber
&+&\left.\frac{s-rs'}{2r^2(r-s)}\right)
\left\{(n-3)+(n-5)\frac{2\epsilon
s}{r^3}\right\}-\frac{s}{2r^3}\left\{(n-3)(n-4)\right.\\\label{9}
&+&\left.(n-5)(n-6)\frac{\epsilon
s}{r^3}\right\}-\frac{2\lambda'\epsilon}{r^4}\left(1-\frac{s}{r}\right)(s-rs')(n-5),
\end{eqnarray}
where prime refers derivative with respect to $r$ and
$\epsilon=(n-3)(n-4)\epsilon_1$ for the sake of notational
simplicity.

\section{Wormhole Solutions}

In order to discuss the wormhole geometry and solutions, there are
several frameworks and strategies used to find unknown functions.
For instance, we have five unknown functions
$\rho(r),~p_r(r),~p_t(r),~\lambda(r)$ and $s(r)$ in the underlying
case. One may choose some kind of equation of state representing
accelerated expansion of the universe, or different forms of energy
density such as energy density of static spherically symmetric
object with non-commutative geometry having Gaussian or Lorentzian
distributions and galactic halo region etc. The traceless
energy-momentum tensor is also used which is related to the Casimir
effect. In order to construct viable wormhole solutions in
$n$-dimensional Einstein-Gauss-Bonnet gravity, we assume different
forms of energy density of non-commutative geometry in the
following.

Nicolini et al. \cite{24} have improved the short distance behavior
of point-like structures in a new conceptual approach based on
coordinate coherent state formalism to noncommutative gravity. In
their method, curvature singularities which appear in general
relativity, can be eliminated. They have demonstrated that black
hole evaporation process should be stopped when a black hole reaches
a minimal mass. This minimal mass, named black hole remnant, is a
result of the existence of a minimal observable length. This
approach, which is the so-called noncommutative geometry inspired
model, via a minimal length caused by averaging noncommutative
coordinate fluctuations cures the curvature singularity in black
holes. In fact, the curvature singularity at the origin of black
holes is substituted for a regular de-Sitter core. Accordingly, the
ultimate phase of the Hawking evaporation as a novel
thermodynamically steady state comprising a non-singular behavior is
concluded.

It must be noted that, generally, it is not required to consider the
length scale of the coordinate non-commutativity to be the same as
the Planck length. Since, the non-commutativity influences appear on
a length scale connected to that region, they can behave as an
adjustable parameter corresponding to that pertinent scale. The
presence of a universal short distance cut-off leads to the effects
such as in quantum field theory, it curves UV divergences while it
cures curvature singularities in general relativity. In the specific
case of the gravity field equations, the only modification occurs at
the level of the energy-momentum tensor, while $G_{\mu\nu}$ is
formally left unchanged. In non-commutative space, the usual
definition of mass density in the form of Dirac delta function does
not hold. So in this space the usual form of the energy density of
the static spherically symmetry smeared and particlelike
gravitational source requires some other forms of distribution.

In view of the above explanations, we are going to discuss wormhole
solutions with the help of two well-known energy distributions such
as Gaussian and Lorentzian in non-commutative scenario. As an
important remark, the essential aspects of the non-commutativity
approach are not specifically sensitive to any of these
distributions of the smearing effects \cite{23F} rather only
distribution parameter is defferent. The Gaussian source has also
been used by Sushkov \cite{24F} to model phantom-energy supported
wormholes, as well as by Nicolini and Spalluci \cite{25F} for the
purpose of modeling the physical effects of short distance
fluctuations of noncommutative coordinates in the study of black
holes. Galactic rotation curves inspired by a non-commutative
geometry background are discussed \cite{26F}. The stability of a
particular class of thin-shell wormholes in noncommutative geometry
is analyzed elsewhere \cite{27F}.

\subsection{Gaussian Distributed Non-commutative Framework}

An intrinsic characteristic of spacetime is the non-commutativity
which plays an effective role in several areas. It is an interesting
consequence of string theory where the coordinates of spacetime
become non-commutative operators on $D$-brane \cite{23}. The
non-commutativity of spacetime can be converted in the commutator,
$[x^\alpha,x^\beta]=i\theta^{\alpha\beta}$, where
$\theta^{\alpha\beta}$ is an anti-symmetric matrix describing
discretization of spacetime and has dimension $(length)^2$. This
discretization process is similar to the discretization of phase
space by Planck constant. Replacing the point-like structures with
smeared objects, the energy density of the particle-like static
spherically symmetric gravitational source having mass $M$ takes the
following form \cite{24}
\begin{equation}\label{10*}
\rho_{_{nc}}=\frac{M}{(4\pi\theta)^{\frac{n-1}{2}}}e^{-\frac{r^2}{4\theta}},
\end{equation}
where $\theta$ is the non-commutative parameter in Gaussian
distribution. The mass M could be a diffused centralized object such
as a wormhole [23]. It is mentioned here that the smearing effect is
achieved by replacing the Gaussian distribution of minimal length
$\sqrt{\theta}$ with the Dirac delta function. In order to construct
$n$-dimensional Einstein Gauss-Bonnet wormhole geometry, we equate
the energy density given in Eq.(\ref{7}) and $\rho_{_{nc}}$ in
(\ref{10*}) yields the following differential equation
\begin{eqnarray}\label{11}
s'=\frac{1}{1+\frac{2s\epsilon}{r^3}}\left[\left(4-n+(7-n)\frac{s\epsilon}{r^3}\right)+\frac{2r^2M}{(n-2)}(4\pi
\theta)^{\frac{1-n}{2}}e^{-\frac{r^2}{4\theta}}\right].
\end{eqnarray}
The solution of this equation is
\begin{eqnarray}\nonumber
s(r)&=&-\frac{1}{2}\left[\frac{r^3}{\epsilon}\pm\frac{r^{\frac{7-n}{2}}\pi^{-\frac{n}{4}}}{\sqrt{\epsilon(n-1)}}\left\{
(n-2)\pi^{\frac{n}{2}}(r^{n-1}+4\epsilon^2
c_1)-4M\sqrt{\pi}\epsilon\right.\right.\\\label{12}&\times&\left.\left.
\textmd{Gamma}[\frac{n-1}{2},\frac{r^2}{4\theta}]\right\}^{\frac{1}{2}}\right],
\end{eqnarray}
where $c_1$ is an integration constant. This solution has two roots
with plus and minus signs and we assign these roots as $s_{+}(r)$
and $s_-(r)$ solutions. In order to plot the quantities
$1-\frac{s}{r},~\rho,~\rho+p_r$ and $\rho+p_t$ to obtain wormhole
solutions for both of these solutions, we restrict ourselves to five
and six-dimensional cases with constant redshift function
$\lambda=0$. The expressions of NEC takes the form
\begin{eqnarray}\label{13}
\rho+p_r&=&\frac{(n-2)}{2r^2}\left(s'-\frac{s}{r}\right)\left(1+\frac{2\epsilon
s}{r^3}\right),\\\label{14}
\rho+p_t&=&\frac{rs'-s}{2r^3}\left(1+\frac{6\epsilon
s}{r^3}\right)+\frac{s}{r^3}\left[(n-3)+(n-5)\frac{2s\epsilon}{r^3}\right].
\end{eqnarray}

In this regard, we assume values of some constants as
$M=0.008,~\theta=0.002$ while $\epsilon_1=0.5,~-0.5$ result
$\epsilon=1,~-1$ for five-dimensional and $\epsilon=3,~-3$ for
six-dimensional case.
\begin{figure} \centering
\epsfig{file=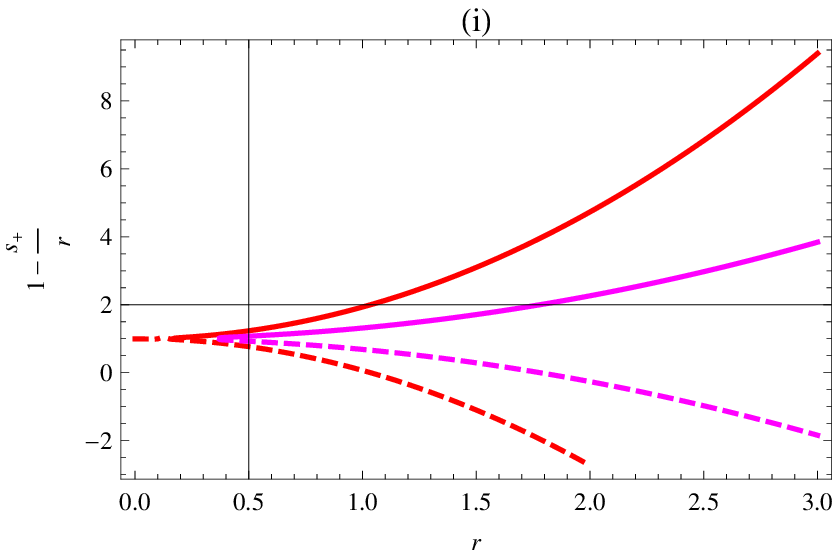,width=.50\linewidth}\epsfig{file=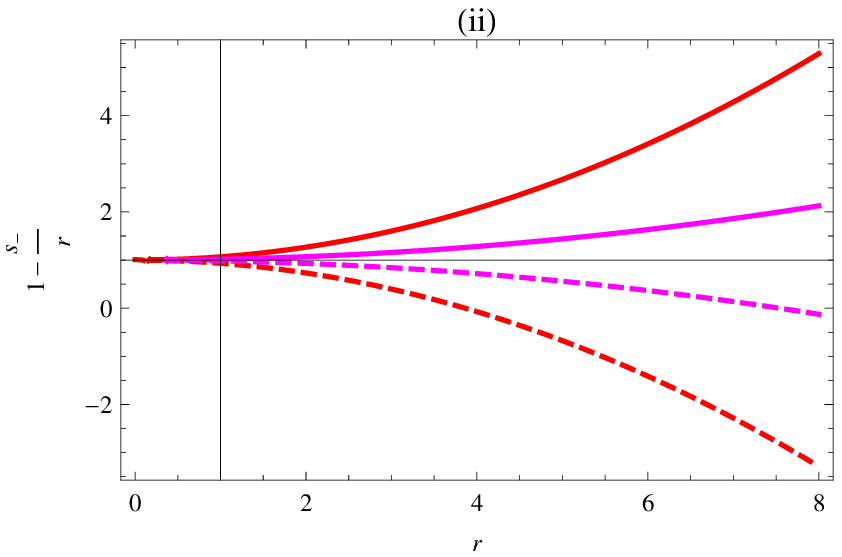,width=.50\linewidth}
\caption{Plots of (i) $1-\frac{s_{+}}{r}$ and (ii)
$1-\frac{s_{-}}{r}$ versus $r$ in Gaussian distribution for
$n=5,~\epsilon=1$ (red), $n=5,~\epsilon=-1$ (red dashed),
$n=6,~\epsilon=3$ (purple) and $n=6,~\epsilon=-3$ (purple dashed).}
\end{figure}
\begin{figure} \centering
\epsfig{file=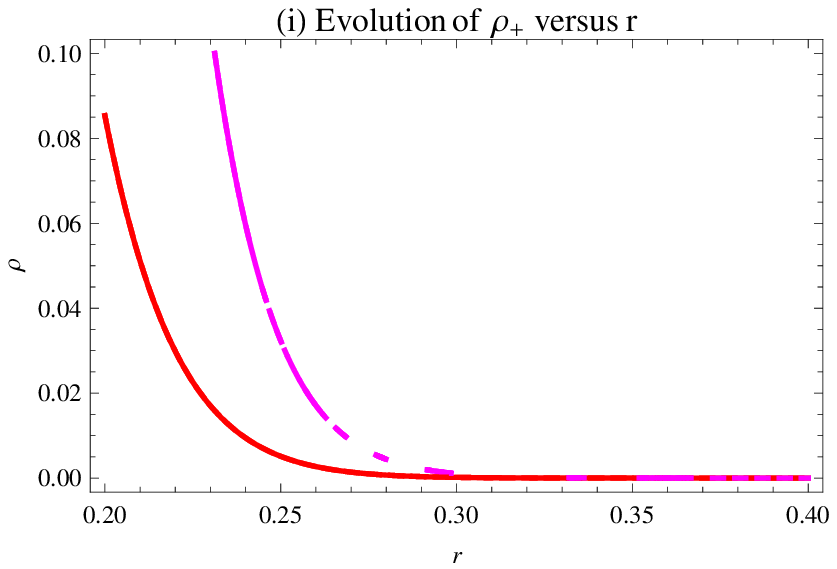,width=.50\linewidth}\epsfig{file=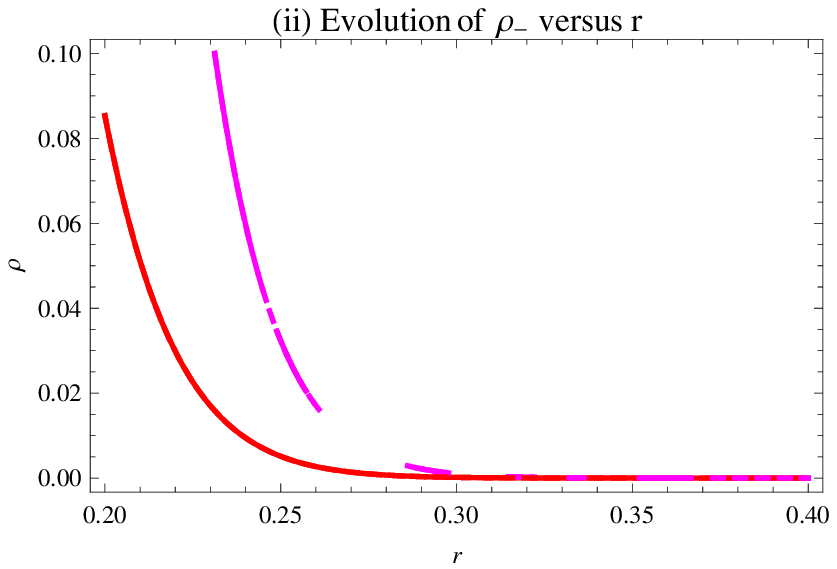,width=.50\linewidth}
\caption{Plots of (i) $\rho_+$ and (ii) $\rho_-$ versus $r$ in
Gaussian distribution for $n=5,~\epsilon=1$ (red),
$n=5,~\epsilon=-1$ (red dashed), $n=6,~\epsilon=3$ (purple) and
$n=6,~\epsilon=-3$ (purple dashed).}
\end{figure}
\begin{figure} \centering
\epsfig{file=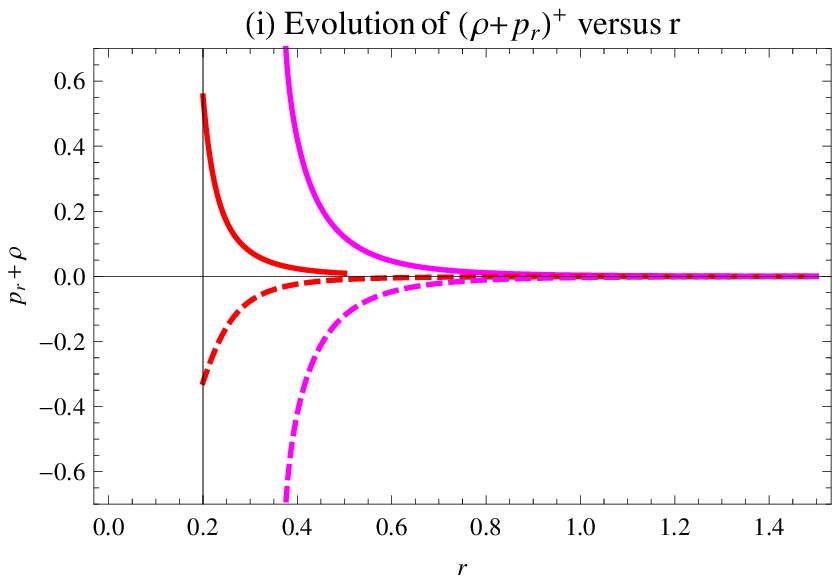,width=.50\linewidth}\epsfig{file=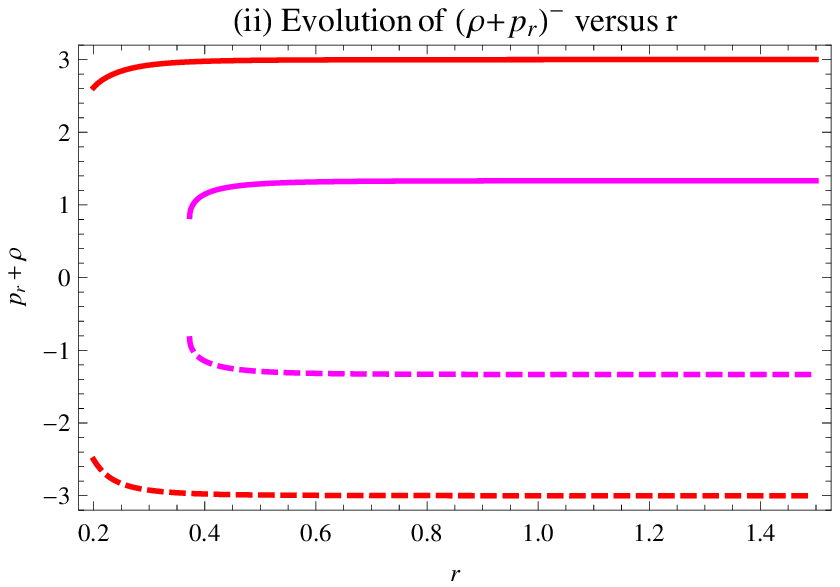,width=.50\linewidth}
\caption{Plots of (i) $\rho+p_r$ for $s_+$ and (ii) $\rho+p_r$ for
$s_-$ versus $r$ in Gaussian distribution for $n=5,~\epsilon=1$
(red), $n=5,~\epsilon=-1$ (red dashed), $n=6,~\epsilon=3$ (purple)
and $n=6,~\epsilon=-3$ (purple dashed).}
\end{figure}
\begin{figure} \centering
\epsfig{file=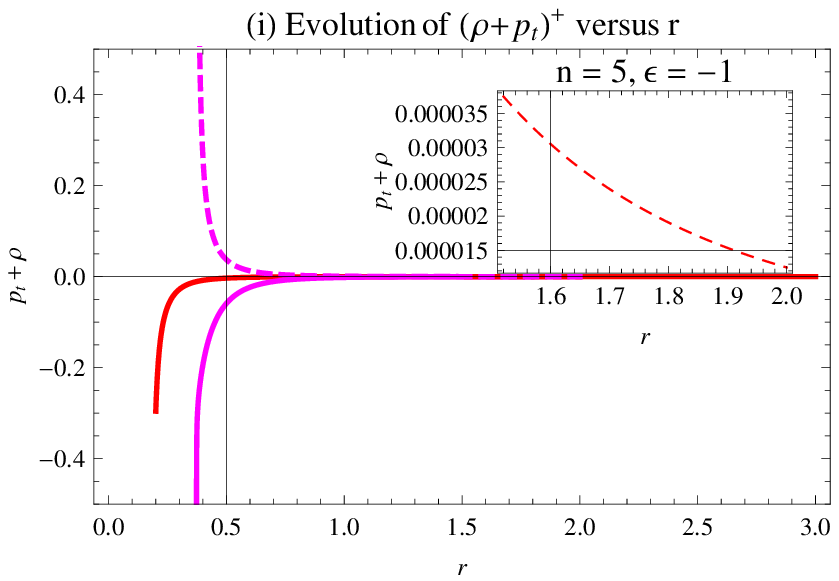,width=.50\linewidth}\epsfig{file=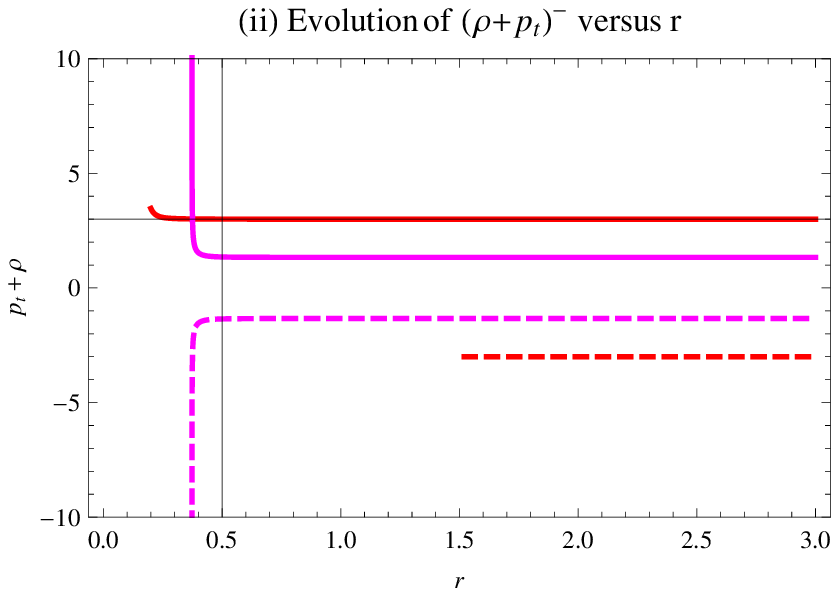,width=.50\linewidth}
\caption{Plots of (i) $\rho+p_t$ for $s_+$ and (ii) $\rho+p_t$ for
$s_-$ versus $r$ for in Gaussian distribution $n=5,~\epsilon=1$
(red), $n=5,~\epsilon=-1$ (red dashed), $n=6,~\epsilon=3$ (purple)
and $n=6,~\epsilon=-3$ (purple dashed).}
\end{figure}
Figure \textbf{1(i)} represents the plot of $1-\frac{s_+}{r}$ versus
$r$ for five and six dimensional wormhole solutions. The graph
represents positively increasing behavior for positive $\epsilon$
for both curves. For negative $\epsilon$, we examine that graph
initially represents positive behavior for both dimensions in the
range $r<1$ and $r<1.8$ for $n=5$ and $6$ respectively, then
decreases towards negative values. In plot \textbf{(ii)}, the graph
of $1-\frac{s_-}{r}$ versus $r$ shows same behavior for all curves
as in plot \textbf{(i)} with positive behavior of dashed curves in
the range $r<4$ and $r<8$ for $n=5$ ad $6$. However, we have less
possibility of wormhole scenario with respect to $r$ for positive
root solution as compared to negative root solution of shape
function.

Figure \textbf{2} shows the behavior of energy density versus $r$ as
positively decreasing behavior for both dimensions corresponding to
$n=5$ and $6$ dimensions. For $s_+$ and positive $\epsilon$, the
plot of $\rho+p_r$ represents positive behavior in decreasing manner
for $n=5$ and $6$ while representing negative behavior for negative
$\epsilon$ as shown in Figure \textbf{3(i)}. In plot \textbf{(ii)},
we examine same behavior for negative root solution. Figure
\textbf{4(i)} depicts the opposite behavior to plot \textbf{3(i)},
i.e., for $n=5,~6$ dimensions and positive $\epsilon,~\rho+p_t$
demonstrates negative behavior. Thus, these plots express the
violation of WEC incorporating the case of $s_+$. For $s_-$, we
obtain same behavior as in plot \textbf{3(ii)} which indicates
positive behavior for positive $\epsilon$ and $n=5,6$ and negative
behavior for negative $\epsilon$ and $n=5,6$. This implies that WEC
satisfies for negative root solution with positive Gauss-Bonnet
coefficient. Thus, we obtain physically acceptable wormhole
solutions satisfying WEC for both dimensions.

\subsection{Lorentzian Distributed Non-commutative Framework}

Now we consider the case of non-commutative geometry with Lorentzian
distribution. The energy density of point-like source under this
distribution becomes \cite{2,11-e}
\begin{equation}\label{10}
\rho_{_{Lnc}}=\frac{M\sqrt{\phi}}{\pi^2(r^2+\phi)^{\frac{n}{2}}},
\end{equation}
where $\phi$ is the non-commutative parameter in Lorentzian
distribution. Inserting $\rho_{_{Lnc}}$ in Eq.(\ref{7}), the
differential equation takes the following form
\begin{eqnarray}\label{15}
s'=\frac{1}{1+\frac{2s\epsilon}{r^3}}\left[\frac{2r^2M\sqrt{\phi}}{\pi^2(n-2)(r^2+\phi)^{\frac{n}{2}}}+
\frac{b}{r}\{4-n+(7-n)\frac{s\epsilon}{r^3}\}\right].
\end{eqnarray}
The solution of this equation is given by
\begin{eqnarray}\nonumber
s(r)&=&\frac{r^3}{2\epsilon\pi}\left[-\pi\pm\frac{1}{(n-2)(n-1)}\left\{\pi^2(n^2-3n+2)(r^n+4r\epsilon^2c_2)
\right.\right.\\\label{16}&+&\left.\left.8M\epsilon
r^n\phi^{\frac{1-n}{2}}\textmd{Hypergeometric2F1}[\frac{1}{2}(n-1),\frac{n}{2},
\frac{n+1}{2},-\frac{r^2}{\phi}]\right\}^{\frac{1}{2}}\right],
\end{eqnarray}
where $c_2$ is an integration constant. We again assign both
solutions as $s_+$ and $s_-$ and explore the wormhole solutions. In
this case, we choose constants as $\phi=2,~M=2,~c_2=0.5$ for same
dimensions and Gauss-Bonnet coefficient as for non-commutative
background.
\begin{figure} \centering
\epsfig{file=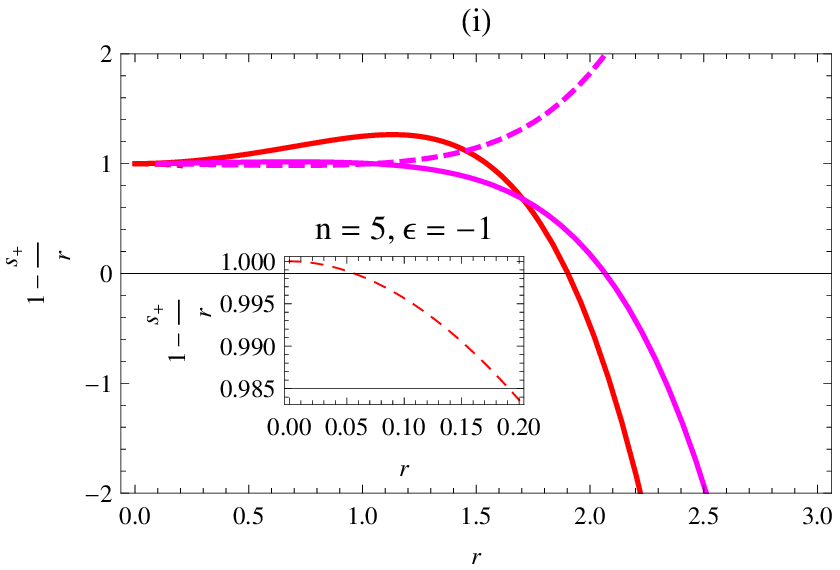,width=.50\linewidth}\epsfig{file=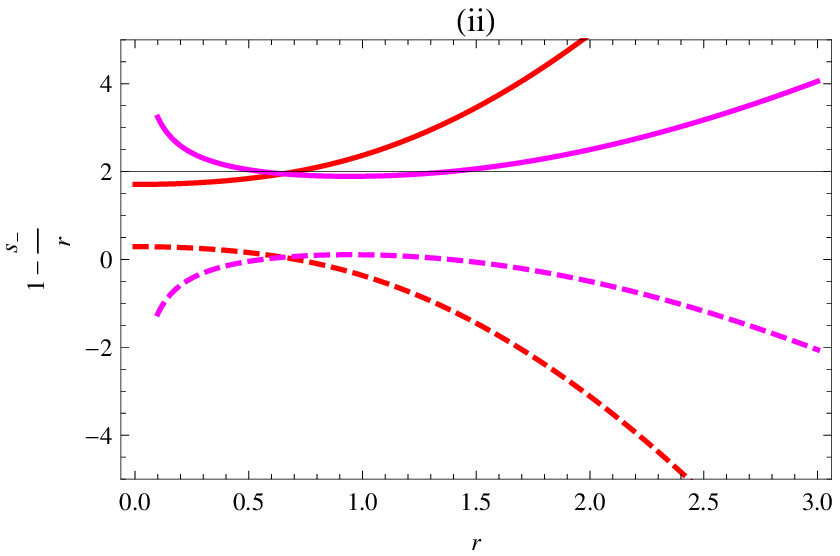,width=.50\linewidth}
\caption{Plots of (i) $1-\frac{s_{+}}{r}$ and (ii)
$1-\frac{s_{-}}{r}$ versus $r$ in Lorentzian distribution for
$n=5,~\epsilon=1$ (red), $n=5,~\epsilon=-1$ (red dashed) and
$n=6,~\epsilon=3$ (purple).}
\end{figure}
\begin{figure} \centering
\epsfig{file=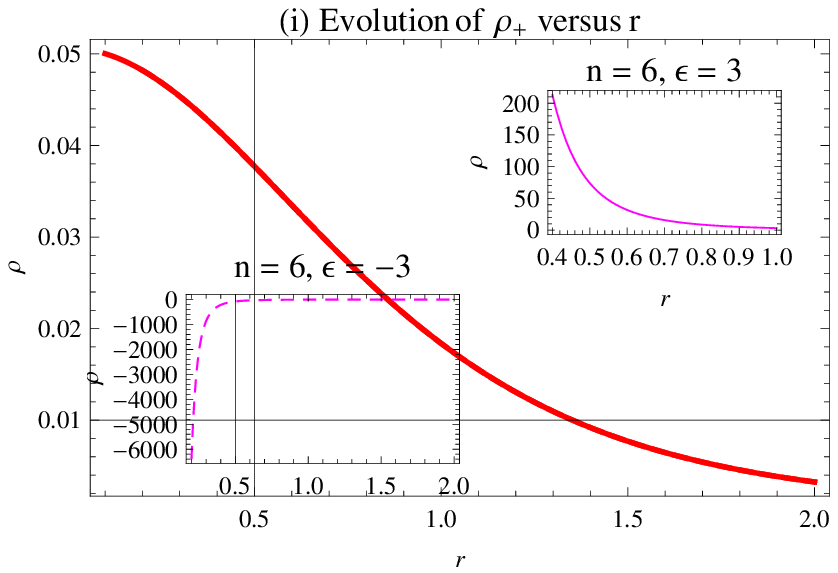,width=.50\linewidth}\epsfig{file=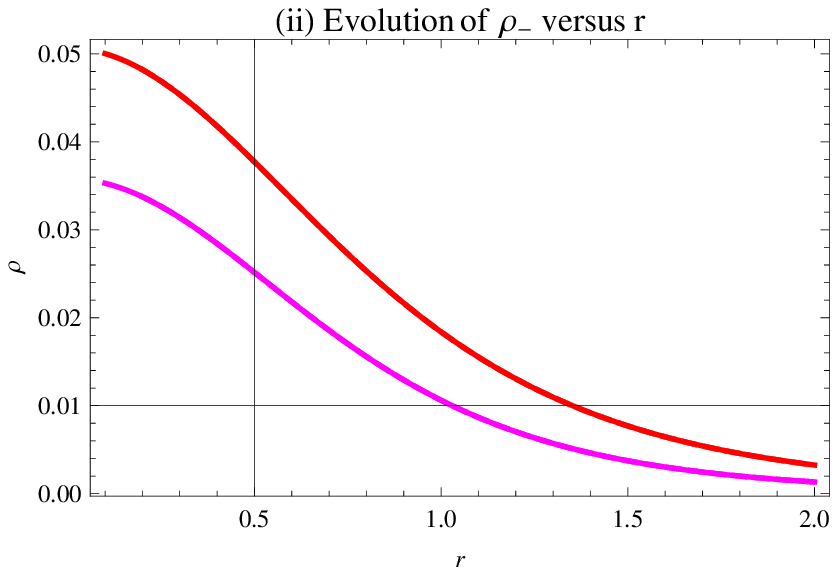,width=.50\linewidth}
\caption{Plots of (i) $\rho_+$ and (ii) $\rho_-$ versus $r$ in
Lorentzian distribution for $n=5,~\epsilon=1$ (red),
$n=5,~\epsilon=-1$ (red dashed) and $n=6,~\epsilon=3$ (purple). For
(i) $\rho_+$, $n=6,~\epsilon=-3$ (purple dashed).}
\end{figure}
\begin{figure} \centering
\epsfig{file=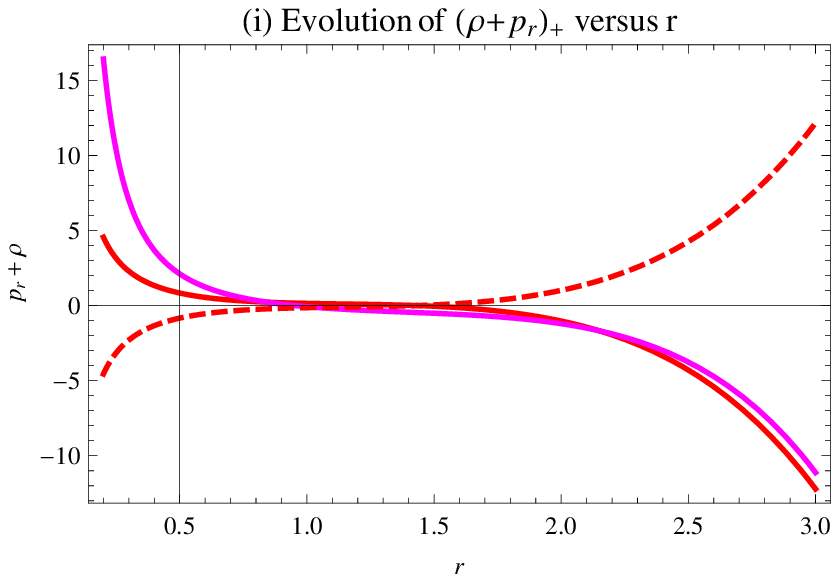,width=.50\linewidth}\epsfig{file=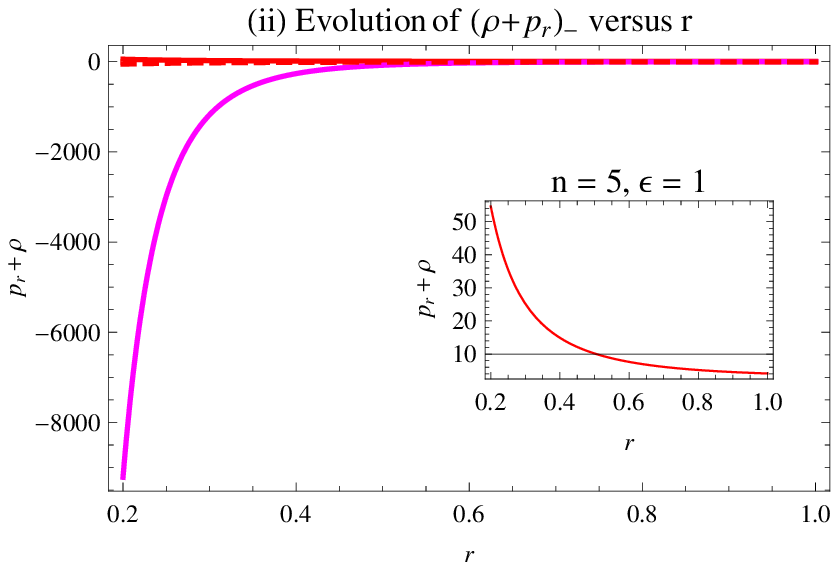,width=.50\linewidth}
\caption{Plots of (i) $\rho+p_r$ for $s_+$ and (ii) $\rho+p_r$ for
$s_-$ versus $r$ in Lorentzian distribution for $n=5,~\epsilon=1$
(red), $n=5,~\epsilon=-1$ (red dashed) and $n=6,~\epsilon=3$
(purple).}
\end{figure}
\begin{figure} \centering
\epsfig{file=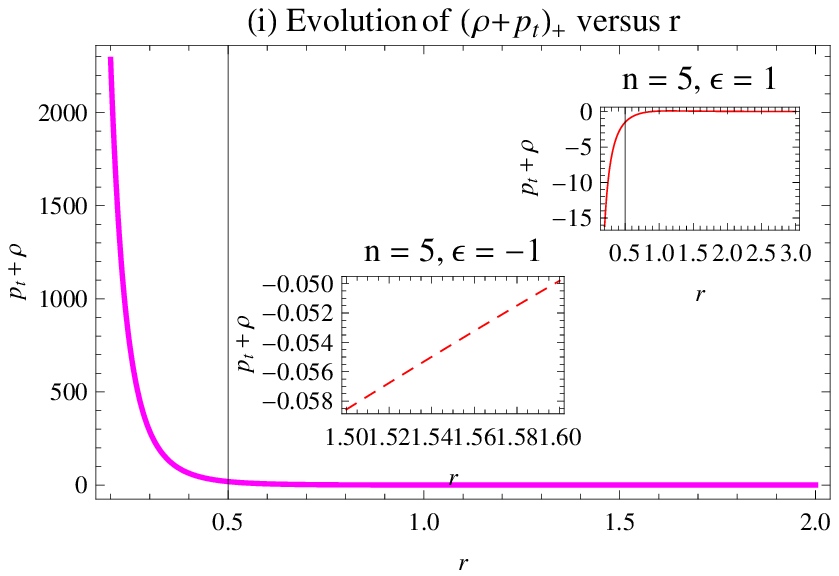,width=.50\linewidth}\epsfig{file=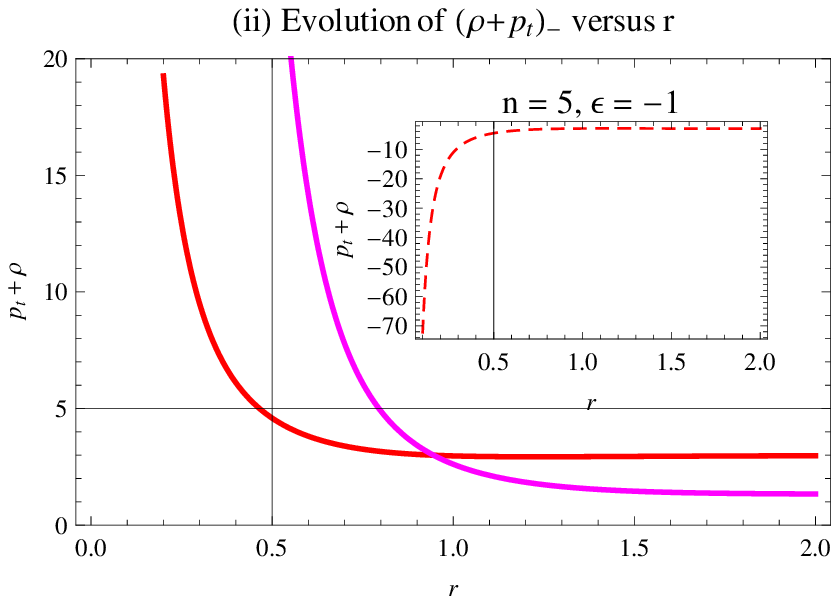,width=.50\linewidth}
\caption{Plots of (i) $\rho+p_t$ for $s_+$ and (ii) $\rho+p_t$ for
$s_-$ versus $r$ in Lorentzian distribution for $n=5,~\epsilon=1$
(red), $n=5,~\epsilon=-1$ (red dashed) and $n=6,~\epsilon=3$
(purple).}
\end{figure}

In Figure \textbf{5(i)}, we plot $1-\frac{s_+}{r}$ versus $r$ which
represents that the condition for wormhole geometry, i.e.,
$1-\frac{s}{r}>0$ holds for $5$ and $6$ dimensions with negative
$\epsilon$. For positive $\epsilon$, the positivity of this
expression depends on some ranges such as, it remains positive for
$r<1.9$ for $n=5$ while $n=6$ observes the range $r<2.1$. The plot
\textbf{5(ii)} corresponds to the plot of $1-\frac{s_-}{r}$ with
respect to $r$ shows positive behavior for positive $\epsilon$. For
$n=5$, it expresses a very short range, $r<0.5$, for positivity
whereas it remains negative for $6$-dimensional solution. That is,
we have no wormhole solution for $n=6,~\epsilon=-3$ taking negative
root solution. So, we skip this case in further discussion. The
behavior of $\rho_+$ and $\rho_-$ remains positive for all cases
except $n=6,~\epsilon=-3$ for which it describes negative behavior,
as shown in Figure \textbf{6(i)-(ii)}. This implies that, we have no
wormhole solutions for $6$-dimensional case with negative
Gauss-Bonnet coefficient for both solutions.

Figure \textbf{7(i)} expresses the behavior of $\rho+p_r$ for
positive root solution versus $r$ which remains positive for the
range $r<1$ for $n=5$ and $6$ with positive $\epsilon$ and then turn
towards negative behavior. It demonstrates negative behavior for
$r<1$ and then moves to positive region. Incorporating $s_-$
solution, $\rho+p_r$ shows positive behavior only for the case
$n=5,~\epsilon>0$ and negative behavior for the remaining two cases
as shown in plot \textbf{7(ii)}. In the Figure \textbf{8(i)-(ii)},
we draw $\rho+p_t$ for both solutions versus $r$. This expression
represents the negative behavior for
$(n=5,~\epsilon=1),(n=5,~\epsilon=-1)$ with $s_+$,
$(n=5,~\epsilon=-1)$ with $s_-$ and preserves positivity for
$(n=6,~\epsilon=3)$ with $s_+,~(n=5,~\epsilon=1),~(n=6,~\epsilon=3)$
with $s_-$ solution.

\section{Equilibrium Condition}

In order to find the equilibrium configuration of the wormhole
solutions in Gaussian as well as Lorentzian distributed
non-commutative backgrounds, we use the generalized
Tolman-Oppenheimer-Volkov equation. This equation is derived by
solving the Einstein equations for a general time-invariant,
spherically symmetric metric having metric tensor
$g_{\alpha\beta}=(e^{\tau(r)},-e^{\omega(r)},-r^2,-r^2\sin^2\theta)$
where $\alpha\beta$ represents only diagonal entries and
$\tau,~\omega$ are general metric functions dependent on $r$. The
generalized Tolman-Oppenheimer-Volkov equation is
\begin{equation}\label{18}
\frac{dp_r}{dr}+\frac{\tau'}{2}(\rho+p_r)+\frac{2}{r}(p_r-p_t)=0.
\end{equation}
Keeping in mind the above equation, Ponce de Le$\acute{o}$n
\cite{EC} proposed an equation for anisotropic mass distribution
which naturally gives the equilibrium for the wormhole subject. It
is given by
\begin{equation}\label{17}
\frac{2}{r}(p_t-p_r)-\frac{e^{\frac{\omega-\tau}{2}}m_{_{eff}}}{r^2}(\rho+p_r)-\frac{dp_r}{dr}=0,
\end{equation}
where effective gravitational mass
$m_{_{eff}}=\frac{1}{2}r^2e^{\frac{\tau-\omega}{2}}\tau'$ is
measured from throat to some arbitrary radius $r$. Accordingly, the
gravitational, hydrostatic as well as anisotropic force due to
anisotropic matter distribution are defied as follows
\begin{equation}\nonumber
f_{_{g}}=-\frac{\tau'(\rho+p_r)}{2},\quad f_{_{h}}=-\frac{dp_r}{dr},
\quad f_{_{a}}=\frac{2(p_t-p_r)}{r}.
\end{equation}
It is required that $f_{_{g}}+f_{_{h}}+f_{_{a}}=0$ must hold for the
wormhole solutions to be in equilibrium.
\begin{figure} \centering
\epsfig{file=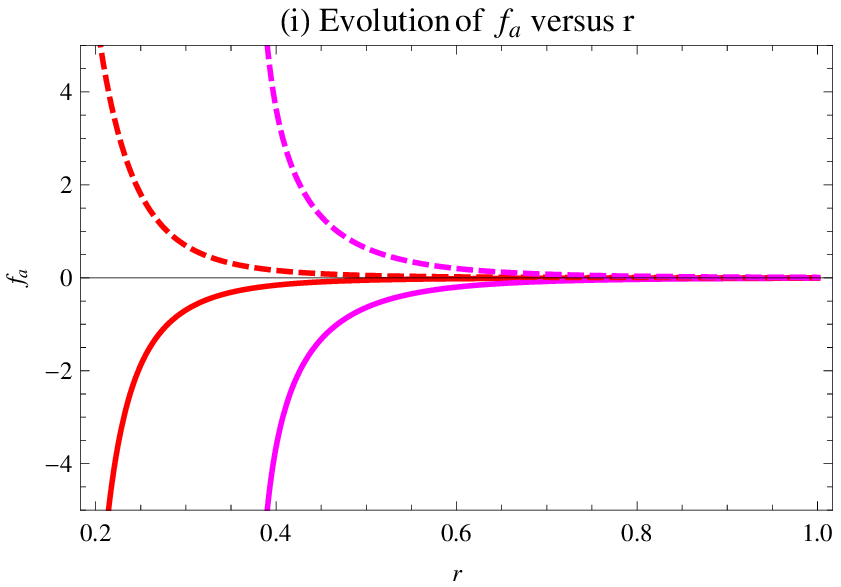,width=.50\linewidth}\epsfig{file=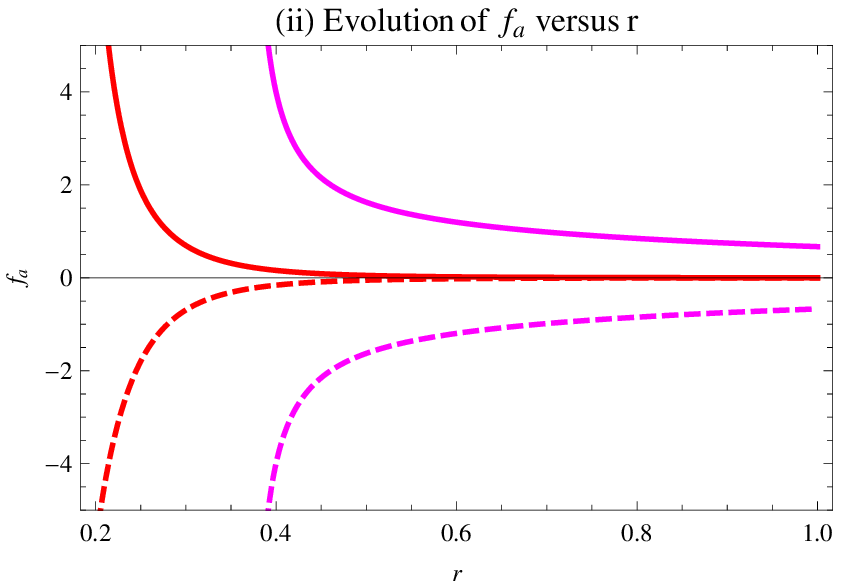,width=.50\linewidth}\\
\epsfig{file=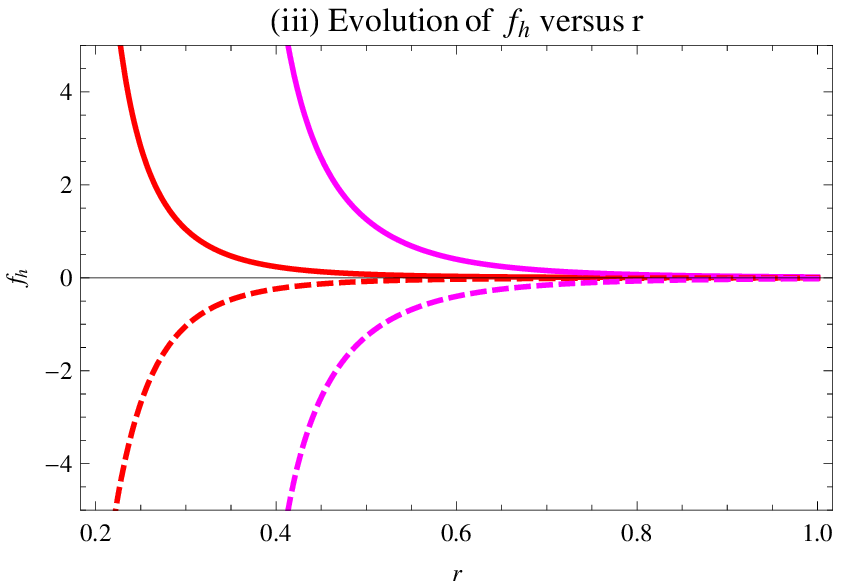,width=.50\linewidth}\epsfig{file=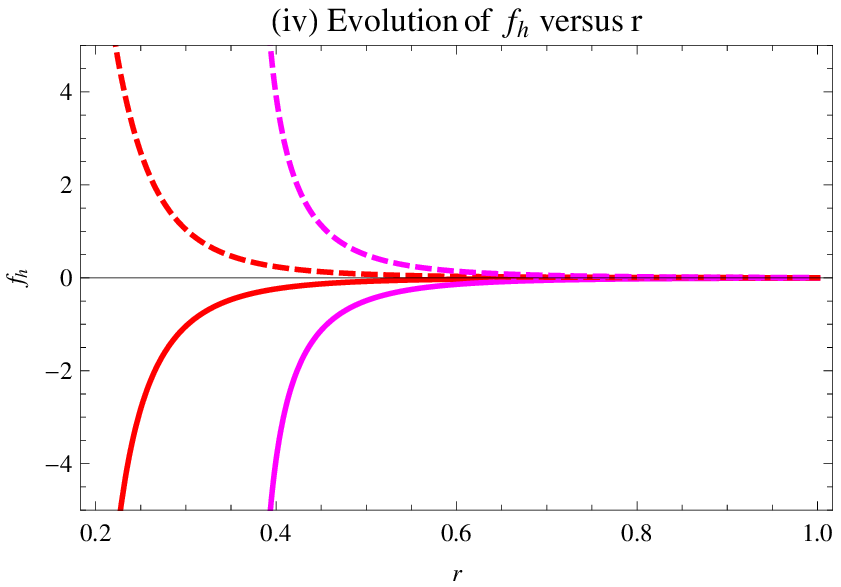,width=.50\linewidth}
\caption{Plots of (i) $f_a$ for $s_+$, (ii) $f_a$ for $s_-$, (iii)
$f_h$ for $s_+$, (iv) $f_h$ for $s_-$ versus $r$ with Gaussian
distribution for $n=5,~\epsilon=1$ (red), $n=5,~\epsilon=-1$ (red
dashed) and $n=6,~\epsilon=3$ (purple).}
\end{figure}
\begin{figure} \centering
\epsfig{file=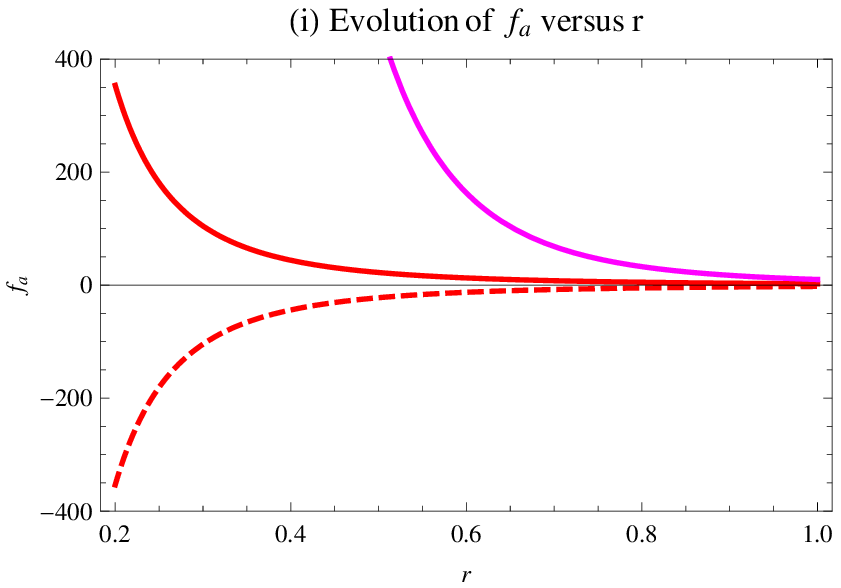,width=.50\linewidth}\epsfig{file=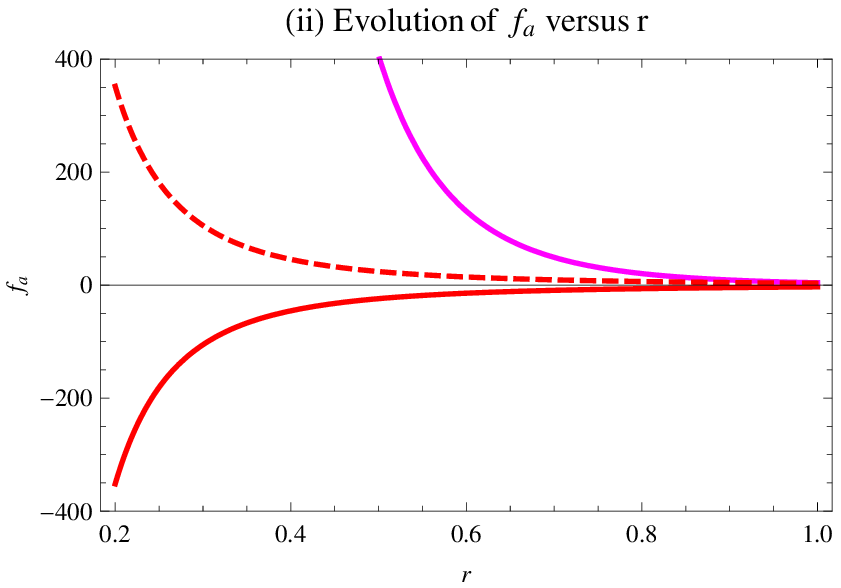,width=.50\linewidth}\\
\epsfig{file=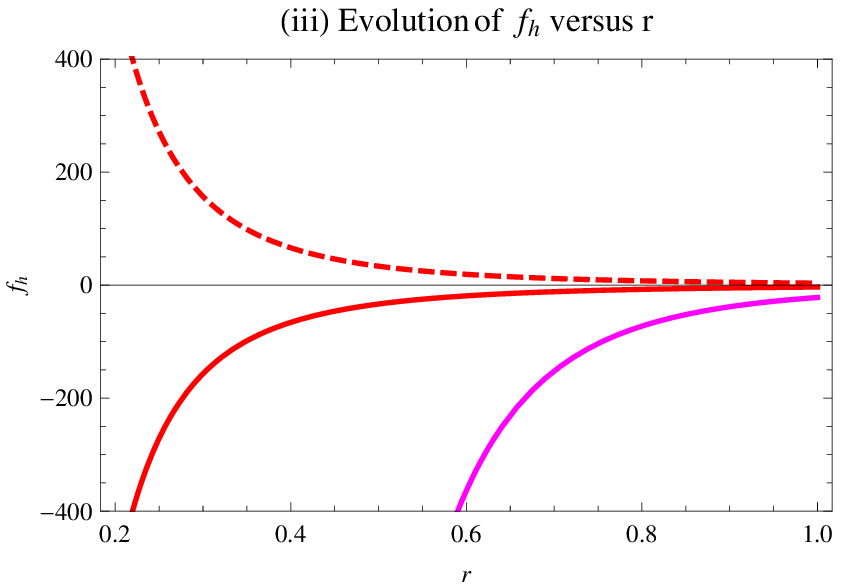,width=.50\linewidth}\epsfig{file=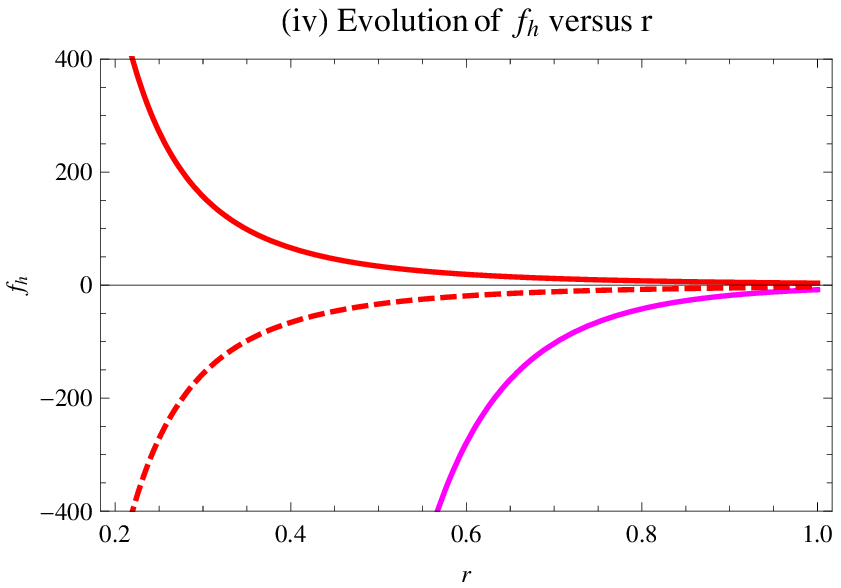,width=.50\linewidth}
\caption{Plots of (i) $f_a$ for $s_+$, (ii) $f_a$ for $s_-$, (iii)
$f_h$ for $s_+$, (iv) $f_h$ for $s_-$ versus $r$ with Lorentzian
distribution for $n=5,~\epsilon=1$ (red), $n=5,~\epsilon=-1$ (red
dashed) and $n=6,~\epsilon=3$ (purple).}
\end{figure}
In the underlying cases, we proceed with constant redshift function
$\lambda=0$ which vanishes the gravitational contribution $f_{_g}$
in the equilibrium equation, i.e., $\tau=2\lambda$ leads to
$\tau'=0$ for constant $\lambda$. Thus, we are left with hydrostatic
and anisotropic forces with corresponding equilibrium condition as
\begin{equation}\label{19}
f_{_{a}}+f_{_h}=0.
\end{equation}

Using Eqs.(\ref{8}) and (\ref{9}), we obtain the following
expressions for the hydrostatic and anisotropic forces
\begin{eqnarray*}
f_{_a}&=&\frac{1}{r^3}\left\{(n-1)\frac{s}{r}-s'\right\}\left(n-3+(n-5)\frac{s\epsilon}{r^3}\right)-\frac{s}{r^4}
\left[(n-3)(n-4)\right.\\&+&\left.(n-5)(n-6)\frac{s\epsilon}{r^3}\right],\\
f_{_h}&=&\frac{(n-2)}{2r^3}\left[n-3+(n-5)\frac{s\epsilon}{r^3}+\frac{s\epsilon(n-5)}{r^3}\right]
\left(s'-\frac{3s}{r}\right).
\end{eqnarray*}
Figure \textbf{9} represents the plots of anisotropic as well as
hydrostatic forces for the wormhole solutions in Gaussian
distributed framework. Plots \textbf{(i)} and \textbf{(iii)}
describe the equilibrium for positive root solution for
$n=5,~\epsilon=1,~-1$ and $n=6,~\epsilon=3,~-3$ through opposite
behavior and hence cancel each other in order to satisfy
Eq.(\ref{19}). For negative root solution, $s_-$, plots
\textbf{(ii)} and \textbf{(iv)} show the stable configuration of
wormhole solutions for fifth dimensional case only. The behavior of
both forces (which are not in opposite manner) does not cancel each
other so no equilibrium configuration examined for sixth dimensional
wormhole solutions. In the case of Lorentzian non-commutative
background, Figure \textbf{10} expresses that all wormhole solutions
are in equilibrium by satisfying the equilibrium condition for both
dimensions.

\section{Conclusion}

It is a well-known fact that the existence of wormhole solutions is
based on violation of NEC. Since the normal matter satisfy the
energy conditions so this violation is associated with an
energy-momentum tensor which provides exotic matter, a hypothetical
form of matter. To explore realistic model or physically acceptable
wormhole solutions, it is necessary to find such a source which
gives the violation of NEC while normal matter meets the energy
conditions. Here in this paper, we have explored wormhole solutions
in $n$-dimensional Einstein Gauss-Bonnet gravity with Gaussian and
Lorentzian non-commutative backgrounds. We have restricted ourselves
to fifth and sixth dimensional cases with positive as well as
negative Gauss-Bonnet coefficient. Also, we have checked the
condition of equilibrium for the wormhole solutions.

The results are summarized in the following tables. It is noted that
the $s_+$ (Table \textbf{1} and \textbf{3}) and $s_-$ (Table
\textbf{2} and \textbf{4}) represent the two roots of the solutions
in both backgrounds and E.C denotes the equilibrium condition.
\begin{table}[h]
\centering \caption{ Wormhole solutions with Gaussian distributed
non-commutative framework for $s_+$.} \vspace{0.5cm}
\begin{small}
\begin{tabular}{|c|c|c|c|c|}
\hline\textbf{$\textmd{Expressions}$}&\textbf{$n=5,\epsilon=1$}&\textbf{$n=5,\epsilon=-1$}&\textbf{$n=6,\epsilon=3$}
&\textbf{$n=6,\epsilon=-3$}\\
\hline\textbf{$1-\frac{s_+}{r}$}&positive&positive for $r<1$&positive&positive for $r<1.8$\\
\hline\textbf{$\rho$}&positive&positive&positive&positive\\
\hline\textbf{$\rho+p_r$}&positive&negative&positive&negative\\
\hline\textbf{$\rho+p_t$}&negative&positive&negative&positive\\
\hline\textbf{$\textmd{WEC}$}&violates&violates&violates&violates\\
\hline\textbf{$\textmd{E.C}$}&holds&holds&holds&holds\\
\hline
\end{tabular}
\end{small}
\end{table}
\begin{table}[bht]
\centering \caption{Wormhole solutions with Gaussian distributed
non-commutative framework for $s_-$.} \vspace{0.5cm}
\begin{small}
\begin{tabular}{|c|c|c|c|c|}
\hline\textbf{$\textmd{Expressions}$}&\textbf{$n=5,\epsilon=1$}&\textbf{$n=5,\epsilon=-1$}&\textbf{$n=6,\epsilon=3$}
&\textbf{$n=6,\epsilon=-3$}\\
\hline\textbf{$1-\frac{s_-}{r}$}&positive&positive for $r<4$&positive&positive for $r<8$\\
\hline\textbf{$\rho$}&positive&positive&positive&positive\\
\hline\textbf{$\rho+p_r$}&positive&negative&positive&negative\\
\hline\textbf{$\rho+p_t$}&positive&negative&positive&negative\\
\hline\textbf{$\textmd{WEC}$}&holds&violates&holds&violates\\
\hline\textbf{$\textmd{E.C}$}&holds&holds& does not hold&does not hold\\
\hline
\end{tabular}
\end{small}
\end{table}
\newpage
In the Lorentzan distributed non-commutative framework, we have
found negative energy density for $s_+$ while violation of
$1-\frac{s}{r}>0$ incorporating $s_-$ for the case
$n=6,~\epsilon=-3$ so we skipped this case. The remaining results
are summarized in the following tables.
\begin{table}[h]
\centering \caption{Wormhole solutions with Lorentzian
non-commutative background for $s_+$.} \vspace{0.5cm}
\begin{small}
\begin{tabular}{|c|c|c|c|}
\hline\textbf{$\textmd{Expressions}$}&\textbf{$n=5,\epsilon=1$}&\textbf{$n=5,\epsilon=-1$}&\textbf{$n=6,\epsilon=3$}\\
\hline\textbf{$1-\frac{s_+}{r}$}&positive for $r<1.9$&positive&positive for $r<2.1$\\
\hline\textbf{$\rho$}&positive&positive&positive\\
\hline\textbf{$\rho+p_r$}&positive for $r<1$&positive for $r>1$&positive for $r<1$\\
\hline\textbf{$\rho+p_t$}&negative&negative&positive\\
\hline\textbf{$\textmd{WEC}$}&does not hold&does not hold&holds for $r<1$\\
\hline\textbf{$\textmd{E.C}$}&holds&holds&holds\\
\hline
\end{tabular}
\end{small}
\end{table}
\begin{table}[h]
\centering \caption{Wormhole solutions with Lorentzian
non-commutative background for $s_-$.} \vspace{0.5cm}
\begin{small}
\begin{tabular}{|c|c|c|c|c|}
\hline\textbf{$\textmd{Expressions}$}&\textbf{$n=5,\epsilon=1$}&\textbf{$n=5,\epsilon=-1$}&\textbf{$n=6,\epsilon=3$}\\
\hline\textbf{$1-\frac{s_-}{r}$}&positive&positive for
$r<0.5$&positive\\
\hline\textbf{$\rho$}&positive&positive&positive\\
\hline\textbf{$\rho+p_r$}&positive&negative&negative\\
\hline\textbf{$\rho+p_t$}&positive&negative&positive\\
\hline\textbf{$\textmd{WEC}$}&holds&does not hold&does not hold\\
\hline\textbf{$\textmd{E.C}$}&holds&holds&holds\\
\hline
\end{tabular}
\end{small}
\end{table}\\

In the paper \cite{8}, higher-dimensional asymptotically flat
wormhole solutions have been explored in the framework of
Gauss-Bonnet gravity by considering a specific choice for a radial
dependent redshift function and by imposing a particular equation of
state. The WEC is satisfied at the throat by considering a negative
Gauss-Bonnet coupling constant. Furthermore, they have considered a
constant redshift function and shown specifically that, for negative
Gauss-Bonnet coupling constant, one may have normal matter in a
determined radial region and that the increase of coupling constant
enlarges the normal matter region. In the present paper, we have
have taken energy density under non-commutative geometry
distributions instead particular equation of state. We have obtained
results for positive Gauss-Bonnet coefficient satisfying energy
conditions. It contains fifth dimensional wormhole solutions in both
backgrounds satisfying equilibrium condition and sixth dimensional
with disequilibrium in non-commutative background with $s_-$
solution. Also, there is possibility for the existence of wormhole
in equilibrium which satisfying WEC for $n=6,\epsilon=3$ taking into
account $s_+$ solution for the range $r<1$.

\end{document}